# ON STORED ENERGIES AND RADIATION Q

Wen Geyi

*Abstract*—This paper discusses the methods for evaluating the stored electromagnetic energies and the radiation Q for an arbitrary lossless antenna. New expressions for the stored electromagnetic energies are derived by using the Poynting theorem in the complex frequency domain, and they are compared with previous theory and are validated by numerical examples. The minimization of radiation Q is also investigated. There exists an optimal current distribution that minimizes the radiation Q for specified antenna geometry. The optimized Q and the optimal current distribution may be determined by solving a generalized eigenvalue equation obtained from the Rayleigh quotient for the radiation Q.

## I. INTRODUCTION

THE antenna (or radiation) Q has been a research topic for many years [1]-[21]. According to the IEEE Standard Definitions of Terms for Antennas, the quality factor of a resonant antenna is defined as the ratio of $2\pi$ times energy stored in the fields excited by the antenna to the energy radiated and dissipated per cycle:

$$Q = \frac{\omega(\tilde{W}_e + \tilde{W}_m)}{P^{rad}}. \qquad (1)$$

The IEEE standard definition (1) applies to an antenna under any conditions, at resonance or above resonance. When the antenna is tuned to resonance, the above definition reduces to the traditional definition for antenna Q [1]. Since the fractional bandwidth is defined at resonant frequency, the result that the fractional bandwidth is the inversion of antenna Q is still valid for the IEEE standard definition(1). The total stored energy $\tilde{W}_e + \tilde{W}_m$ in (1) may be defined as the difference between the total energy and the total radiated energy produced by the antenna:

$$\tilde{W}_e + \tilde{W}_m = W_e + W_m - \frac{r}{v}P^{rad}, \ r \to \infty. \qquad (2)$$

where $W_e$ and $W_m$ are the total electric field and magnetic field energy in the whole space respectively, $r$ is the radius of the sphere that encloses the antenna, and $v = 1/\sqrt{\mu\varepsilon}$. The definition was first proposed by Counter [2] and has been widely adopted by antenna society [3]-[5]. Note that the total stored energy (2) has taken account of all the stored energy around the antenna, including the part of the energy inside the circumscribing sphere of the antenna [5]. Equation (2) has been directly used to calculate the stored energy of antennas in [6] and [7] by using FDTD. The stored energy quickly becomes stable when $r$ is increased to one or two wavelengths.

Figure 1 shows the first antenna model consisting of an impressed source (region 1), a feeding line (region 2) and a scatterer (antenna) connected to the feeding line (region 3). The impressed source $\mathbf{J}_{imp}$ induces the current $\mathbf{J}$ on the antenna (i.e., the scatterer) through the feeding line. If the antenna system is assumed to be lossless, the stored energies can be expressed as [5]

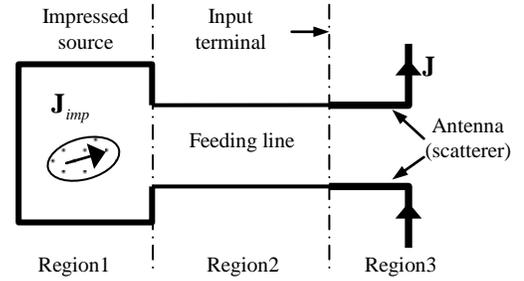

Figure 1 First antenna model: an arbitrary antenna fed by a waveguide.

$$\tilde{W}_e = \frac{1}{8}|I|^2 \left( \frac{\partial X}{\partial \omega} - \frac{X}{\omega} \right), \qquad (3)$$

$$\tilde{W}_m = \frac{1}{8}|I|^2 \left( \frac{\partial X}{\partial \omega} + \frac{X}{\omega} \right), \qquad (4)$$

where $I$ is antenna input terminal current and $X$ is the antenna input reactance defined by

$$X = \frac{4\omega}{|I|^2}(\tilde{W}_m - \tilde{W}_e) \qquad (5)$$

Equations (3) and (4) are the Foster reactance theorem for a lossless antenna. The energy difference $\tilde{W}_m - \tilde{W}_e$ and the radiated power $P^{rad}$ can be determined by the source distributions. From the Poynting theorem in frequency domain, we may obtain [8]

$$P^{rad} = \frac{\omega\eta v}{8\pi} \cdot \iint_{V\ V} \left[ \frac{1}{v^2} \frac{\bar{\mathbf{J}}(\mathbf{r}) \cdot \mathbf{J}(\mathbf{r}')}{R} - \frac{\bar{\rho}(\mathbf{r})\rho(\mathbf{r}')}{R} \right] \sin(kR) dV(\mathbf{r}) dV(\mathbf{r}'), \qquad (6)$$

Manuscript received March 20, 2013. This work is supported in part by the Priority Academic Program Development of Jiangsu Higher Education Institutions, in part by a grant from Huawei Technologies Co. Ltd.

W. Geyi is with the College of Electronic and Information Engineering, Nanjing University of Information Science and Technology, Nanjing, 210044, P. R. China, (e-mail: wgy@nuist.edu.cn).

$$\tilde{W}_m - \tilde{W}_e = \frac{\eta v}{16\pi} \cdot$$
$$\iint_{VV} \left[ \frac{1}{v^2} \frac{\overline{\mathbf{J}}(\mathbf{r}) \cdot \mathbf{J}(\mathbf{r}')}{R} - \frac{\overline{\rho}(\mathbf{r})\rho(\mathbf{r}')}{R} \right] \cos(kR) dV(\mathbf{r}) dV(\mathbf{r}'), \quad (7)$$

where $V$ stands for the source region that the induced current $\mathbf{J}$ occupies and a bar over a letter designates the complex conjugate operation. Once the current distribution $\mathbf{J}$ and the input terminal current $I$ are known, the stored energies and thus the Q can be determined from (1), (3), (4), (5), (6) and (7).

It is noted that the Foster reactance theorem (3) and (4) for antenna has been a controversial topic for a long time [10]-[13]. Equations (3) and (4) was used by Harrington to study antenna Q and bandwidth in 1968 in his classic book [9] although a rigorous proof of Foster reactance theorem for a radiating system was not available at the time. Harrington believed that (3) and (4) are approximately valid for a high Q network in the vicinity of resonance. In a discussion with Collin [10][11], Rhodes also believed that the fact that the slope of the input reactance can be negative at some frequencies is immaterial; it is always positive at the only frequency (resonance) for which bandwidth and Q are defined. For further discussion, please see the appendix.

Figure 2 shows the second antenna model consisting of a current distribution $\mathbf{J}$, which is confined in a region $V$ bounded by $S$. If the antenna is assumed to be small, the stored energies around the antenna can be determined by [9]

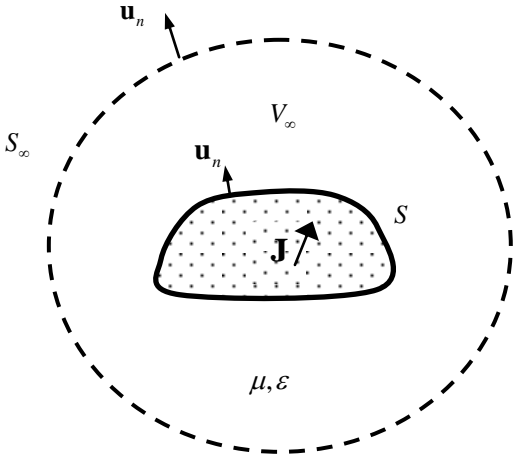

Figure 2 Second antenna model: a scatterer excited by a nearby impressed current.

$$\tilde{W}_e = \frac{\eta v}{16\pi} \iint_{VV} \frac{1}{R} \rho(\mathbf{r})\overline{\rho}(\mathbf{r}') dV(\mathbf{r}) dV(\mathbf{r}'), \quad (8)$$

$$\tilde{W}_m = \frac{\eta v}{16\pi} \frac{1}{v^2} \iint_{VV} \frac{\mathbf{J}(\mathbf{r}) \cdot \overline{\mathbf{J}}(\mathbf{r}')}{R} dV(\mathbf{r}) dV(\mathbf{r}')$$
$$+ \frac{\eta v}{16\pi} \frac{k^2}{2} \iint_{VV} R \rho(\mathbf{r})\overline{\rho}(\mathbf{r}') dV(\mathbf{r}) dV(\mathbf{r}'). \quad (9)$$

The expressions (8) and (9) are obtained by combining the Poynting theorems in both frequency domain and time domain. It has been demonstrated that (8) and (9) give the exact same values as (3) and (4) when antenna is small [5][14], and they have also been verified by other numerical methods [6],[7].

In a recent paper, Vandenbosch has tried to extend (8) and (9) to an antenna of arbitrary size [16]. By assuming that the current is independent of frequency (a similar assumption appears in [15], where the antenna terminal current is assumed to be independent of frequency. This is equivalent to assuming that the aperture field at the antenna input terminal is independent of frequency. Rhodes pointed out that any frequency dependent solution obtained by assuming that the aperture fields are independent of frequency would be totally unrealistic because it would be non-causal [11]. In fact, a quantity in frequency domain essentially represents the Fourier transform of the corresponding quantity in time domain. Therefore a quantity that is constant in a frequency band implies that it must extend from $t=-\infty$ to $t=+\infty$ in time domain, resulting in a causality problem), he has found the following expressions for the stored energies

$$\tilde{W}_e = \frac{\eta v}{16\pi} \iint_{VV} \overline{\rho}(\mathbf{r})\rho(\mathbf{r}') \frac{\cos kR}{R} dV(\mathbf{r}) dV(\mathbf{r}')$$
$$+ \frac{\omega \eta}{32\pi} \iint_{VV} \left[ \overline{\rho}(\mathbf{r})\rho(\mathbf{r}') - \frac{1}{v^2} \overline{\mathbf{J}}(\mathbf{r}) \cdot \mathbf{J}(\mathbf{r}') \right] \sin kR \, dV(\mathbf{r}) dV(\mathbf{r}'), \quad (10)$$

$$\tilde{W}_m = \frac{\eta v}{16\pi} \iint_{VV} \frac{1}{v^2} \overline{\mathbf{J}}(\mathbf{r}) \cdot \mathbf{J}(\mathbf{r}') \frac{\cos kR}{R} dV(\mathbf{r}) dV(\mathbf{r}')$$
$$+ \frac{\omega \eta}{32\pi} \iint_{VV} \left[ \overline{\rho}(\mathbf{r})\rho(\mathbf{r}') - \frac{1}{v^2} \overline{\mathbf{J}}(\mathbf{r}) \cdot \mathbf{J}(\mathbf{r}') \right] \sin kR \, dV(\mathbf{r}) dV(\mathbf{r}'). \quad (11)$$

Vandenbosch's derivation is lengthy and rather involved. It is worth mentioning that (10) and (11) can also be obtained directly from the Foster reactance theorem (3) and (4) in a simple manner. In fact, it follows from (5) that

$$\frac{\partial X}{\partial \omega} = \frac{4(\tilde{W}_m - \tilde{W}_e)}{|I|^2} + \omega \frac{\partial}{\partial \omega} \left[ \frac{4(\tilde{W}_m - \tilde{W}_e)}{|I|^2} \right]. \quad (12)$$

Considering (7) and applying the assumption that the normalized current distribution $\mathbf{J}(r)/I$ is independent of frequency, we have

$$\frac{\partial}{\partial \omega} \frac{\tilde{W}_m - \tilde{W}_e}{|I|^2} = \frac{\eta v}{8\pi} \iint_{V_0 V_0} \frac{\nabla \cdot \overline{\mathbf{J}}(\mathbf{r}) \nabla \cdot \mathbf{J}(\mathbf{r}')}{\omega^3 |I|^2 R} \cos(kR) dV(\mathbf{r}) dV(\mathbf{r}')$$
$$- \frac{\eta}{16\pi} \iint_{V_0 V_0} \left[ \frac{\overline{\mathbf{J}}(\mathbf{r}) \cdot \mathbf{J}(\mathbf{r}')}{v^2 |I|^2} - \frac{\nabla \cdot \overline{\mathbf{J}}(\mathbf{r}) \nabla \cdot \mathbf{J}(\mathbf{r}')}{\omega^2 |I|^2} \right] \sin(kR) dV(\mathbf{r}) dV(\mathbf{r}'). \quad (13)$$

Taking (13) into account and substituting (12) into (3) and (4), we immediately get (10) and (11). Therefore the Foster reactance theorem (3) and (4) actually implies (10) and (11) if the current distribution is assumed to be independent of frequency. A time-domain approach for the calculation of stored energies have also been proposed by Vandenbosch recently [17],[18]. It is noted that the time-domain results for the stored energies do not reduce to the corresponding results in frequency domain [i.e., (10) and (11)] when the fields become sinusoidal [18].

The assumption that the current distribution $\mathbf{J}$ or antenna terminal current does not depend on frequency is, of course,

invalid in most situations. In this paper, a rigorous approach to the stored energies is presented by using the complex frequency domain method. The paper is organized as follows. In Section 2, the complex frequency domain approach is adopted to derive new expressions for the stored energies based on the antenna model shown in Figure 2, in an attempt to generalize the results (8) and (9) for small antenna to an antenna of arbitrary size, without using the above inappropriate assumption. From the Maxwell equations in the complex frequency domain with complex frequency $s = \alpha + j\omega$, the Poynting theorem, a balance equation on the complex power flow in complex frequency domain, can be established. Both sides of the Poynting theorem can be expanded as a Taylor series in the neighbourhood of $\alpha = 0$. By comparing the expansion coefficients of the similar terms and using Cauchy-Riemann conditions, the stored energies can then be determined. Since both the stored energies and radiation power are uniquely determined by the current $\mathbf{J}$, the radiation Q can be considered as a functional of $\mathbf{J}$ and may be expressed as a Rayleigh quotient. In Section 3, the minimization of radiation Q for specified antenna geometry is discussed. The new expressions of the stored energies are compared with the Foster reactance theorem (3) and (4), and they are validated by numerical examples in Section 4. The optimal current distributions that render Q a minimum for several antenna geometries are also presented.

## II. NEW EXPRESSIONS FOR STORED ENERGIES

In order to determine the stored energies $\tilde{W}_e$ and $\tilde{W}_m$, two independent equations are needed. The Poynting theorem in frequency domain provides an equation for the stored electric and magnetic energy. The Poynting theorem in time domain can be used as another independent equation and this strategy has been successfully applied to the study of small antennas [9]. In this section, we demonstrate that the Poynting theorem in complex frequency domain offers two independent equations which can be used to determine the stored energies $\tilde{W}_e$ and $\tilde{W}_m$. Let us consider an arbitrary current distribution $\mathbf{J}$, which occupies a finite region $V$ bounded by $S$ as shown in Figure 2. The current distribution produces electric field $\mathbf{E}$ and magnetic field $\mathbf{H}$. Let $S_\infty$ be a closed surface large enough to enclose the source region $V$. We introduce the complex frequency $s = \alpha + j\omega$ and all calculations are confined to the complex frequency plane. For clarity, all quantities in the complex frequency domain are denoted by the same symbols in real frequency domain but explicitly showing the dependence on $s$. Taking the Laplace transform of the time-domain Maxwell's equations for the fields in a lossless medium yields

$$\nabla \times \mathbf{E}(\mathbf{r},s) = -s\mu \mathbf{H}(\mathbf{r},s),$$
$$\nabla \times \mathbf{H}(\mathbf{r},s) = \mathbf{J}(\mathbf{r},s) + s\varepsilon \mathbf{E}(\mathbf{r},s). \quad (14)$$

The frequency-domain quantities can be recovered by letting $\alpha = 0$ in (14). It follows from (14) that

$$\nabla \cdot \left[\frac{1}{2}\mathbf{E}(\mathbf{r},s) \times \bar{\mathbf{H}}(\mathbf{r},s)\right] = -\frac{1}{2}\mathbf{E}(\mathbf{r},s) \cdot \bar{\mathbf{J}}(\mathbf{r},s)$$
$$-\frac{1}{2}\alpha\left[\mu|\mathbf{H}(\mathbf{r},s)|^2 + \varepsilon|\mathbf{E}(\mathbf{r},s)|^2\right] \quad (15)$$
$$-\frac{1}{2}j\omega\left[\mu|\mathbf{H}(\mathbf{r},s)|^2 - \varepsilon|\mathbf{E}(\mathbf{r},s)|^2\right].$$

Taking the integration of (15) over the region $V_\infty$ bounded by $S_\infty$, as shown in Figure 2, gives

$$\int_{S_\infty}\frac{1}{2}\left[\mathbf{E}(\mathbf{r},s)\times\bar{\mathbf{H}}(\mathbf{r},s)\right]\cdot\mathbf{u}_n dS(\mathbf{r}) + \int_V \frac{1}{2}\mathbf{E}(\mathbf{r},s)\cdot\bar{\mathbf{J}}(\mathbf{r},s)dV(\mathbf{r})$$
$$= -2\alpha[W_m(s) + W_e(s)] - 2j\omega[W_m(s) - W_e(s)], \quad (16)$$

where

$$W_m(s) = \frac{1}{4}\int_{V_\infty}\mu|\mathbf{H}(\mathbf{r},s)|^2 dV(\mathbf{r}), \ W_e(s) = \frac{1}{4}\int_{V_\infty}\varepsilon|\mathbf{E}(\mathbf{r},s)|^2 dV(\mathbf{r}).$$

Letting

$$P^{rad}(s) = \frac{1}{2}\int_{S_\infty}\left[\mathbf{E}(\mathbf{r},s)\times\bar{\mathbf{H}}(\mathbf{r},s)\right]\cdot\mathbf{u}_n dS(\mathbf{r})$$

and substituting it into (16), we get

$$-\int_V \frac{1}{2}\mathbf{E}(\mathbf{r},s)\cdot\bar{\mathbf{J}}(\mathbf{r},s)dV(\mathbf{r}) = P^{rad}(s)$$
$$+2\alpha[W_m(s)+W_e(s)]+2j\omega[W_m(s)-W_e(s)]. \quad (17)$$

In the complex frequency domain, the fields produced by the current $\mathbf{J}(\mathbf{r},s)$ can be represented by

$$\mathbf{E}(\mathbf{r},s) = -\eta v \nabla\int_V \frac{\rho(\mathbf{r}',s)}{4\pi R}e^{-sR/v}dV(\mathbf{r}') - \frac{\eta}{v}s\int_V \frac{\mathbf{J}(\mathbf{r}',s)}{4\pi R}e^{-sR/v}dV(\mathbf{r}'). \quad (18)$$

If $\alpha$ is sufficiently small so that $\alpha \ll v/r$, one can make a first order approximation $e^{-\alpha r/v} \approx 1 - \alpha r/v$, and derive directly from the Maxwell equations, defined in the complex frequency plane, the following:

$$\mathbf{E}^{rad}(\mathbf{r},s) \approx -\left(1-r\frac{\alpha}{v}\right)\frac{j\omega\mu}{4\pi r}e^{-jkr}\cdot$$
$$\int_V\left[\mathbf{J}(\mathbf{r}',s) - \mathbf{J}(\mathbf{r}',s)\cdot\mathbf{u}_r\right]e^{-jk\mathbf{u}_r\cdot\mathbf{r}'}dV(\mathbf{r}').$$

This gives

$$P^{rad}(s) = P^{rad}(\omega)(1 - r\alpha/v)^2 \approx (1 - 2r\alpha/v)P^{rad}(\omega), \quad (19)$$

where $P^{rad}(\omega)$ is the radiated power in the frequency domain, which is independent of $\alpha$. Substituting (19) into (17), we obtain

$$-\int_V \frac{1}{2}\bar{\mathbf{J}}(\mathbf{r},s)\cdot\mathbf{E}(\mathbf{r},s)dS(\mathbf{r}) = P^{rad}(\omega)$$
$$+2\alpha\left[W_m(s)+W_e(s)-\frac{r}{v}P^{rad}(\omega)\right] \quad (20)$$
$$+2j\omega[W_m(s)-W_e(s)].$$

Note that the term in the square bracket on the right-hand side gives the total stored energy at $\alpha = 0$. Introducing the stored energies in the complex frequency domain [1]

$$\tilde{W}_m(s) + \tilde{W}_e(s) = W_m(s) + W_e(s) - \frac{r}{v}P^{rad}(\omega),$$

$$\tilde{W}_m(s) - \tilde{W}_e(s) = W_m(s) - W_e(s),$$

we have

$$-\int_V \frac{1}{2}\overline{\mathbf{J}}(\mathbf{r},s)\cdot \mathbf{E}(\mathbf{r},s)dV(\mathbf{r}) = P^{rad}(\omega)$$
$$+2\alpha\left[\tilde{W}_m(s)+\tilde{W}_e(s)\right]+2j\omega\left[\tilde{W}_m(s)-\tilde{W}_e(s)\right]. \quad (21)$$

Substituting (18) into (21), we obtain

$$-\frac{1}{2}\int_V \overline{\mathbf{J}}(\mathbf{r},s)\cdot \mathbf{E}(\mathbf{r},s)dV(\mathbf{r}) =$$
$$\eta v\overline{s}\iint_{VV}\frac{\overline{\rho}(\mathbf{r},s)\rho(\mathbf{r}',s)}{8\pi R}e^{-sR/v}dV(\mathbf{r})dV(\mathbf{r}') \quad (22)$$
$$+\frac{\eta}{v}s\iint_{VV}\frac{\overline{\mathbf{J}}(\mathbf{r},s)\cdot \mathbf{J}(\mathbf{r}',s)}{8\pi R}e^{-sR/v}dV(\mathbf{r})dV(\mathbf{r}').$$

It follows from (21) and (22) that

$$\frac{\eta v}{16\pi}\iint_{VV}\frac{\overline{\rho}(\mathbf{r},s)\rho(\mathbf{r}',s)}{R}\overline{s}e^{-sR/v}dV(\mathbf{r})dV(\mathbf{r}')$$
$$+\frac{\eta v}{16\pi}\iint_{VV}\frac{1}{v^2}\frac{\overline{\mathbf{J}}(\mathbf{r},s)\cdot \mathbf{J}(\mathbf{r}',s)}{R}se^{-sR/v}dV(\mathbf{r})dV(\mathbf{r}') \quad (23)$$
$$=\frac{1}{2}P^{rad}(\omega)+\alpha\left[\tilde{W}_m(s)+\tilde{W}_e(s)\right]+j\omega\left[\tilde{W}_m(s)-\tilde{W}_e(s)\right].$$

For arbitrary analytic functions $a(s)=a_r(\alpha,\omega)+ja_i(\alpha,\omega)$ and $b(s)=b_r(\alpha,\omega)+jb_i(\alpha,\omega)$, the Cauchy-Riemann conditions hold

$$\frac{\partial a_r(\alpha,\omega)}{\partial \alpha}=\frac{\partial a_i(\alpha,\omega)}{\partial \omega},\quad \frac{\partial a_i(\alpha,\omega)}{\partial \alpha}=-\frac{\partial a_r(\alpha,\omega)}{\partial \omega},$$
$$\frac{\partial b_r(\alpha,\omega)}{\partial \alpha}=\frac{\partial b_i(\alpha,\omega)}{\partial \omega},\quad \frac{\partial b_i(\alpha,\omega)}{\partial \alpha}=-\frac{\partial b_r(\alpha,\omega)}{\partial \omega}. \quad (24)$$

These relations imply

$$\frac{\partial \overline{a}(s)}{\partial \alpha}=j\frac{\partial \overline{a}(s)}{\partial \omega},\quad \frac{\partial a(s)}{\partial \alpha}=j\frac{\partial a(s)}{\partial \omega}.$$

The function $\overline{a}(s)b(s)$ may be expanded into a Taylor series at $\alpha=0$:

$$\overline{a}(s)b(s)\approx \overline{a}(j\omega)b(j\omega)+\alpha[\Omega_1(j\omega)+j\Omega_2(j\omega)]+o(\alpha),$$

where the Cauchy-Riemann conditions (24) have been used, and

$$\Omega_1(j\omega)=\operatorname{Re}\left[j\frac{\partial \overline{a}(j\omega)}{\partial \omega}b(j\omega)-j\overline{a}(j\omega)\frac{\partial b(j\omega)}{\partial \omega}\right],$$
$$\Omega_2(j\omega)=\operatorname{Im}\left[j\frac{\partial \overline{a}(j\omega)}{\partial \omega}b(j\omega)-j\overline{a}(j\omega)\frac{\partial b(j\omega)}{\partial \omega}\right].$$

For small $\alpha$, we may use the following expansions

$$e^{-sR/v}\approx (\cos kR - j\sin kR)-\alpha\frac{R}{v}(\cos kR - j\sin kR).$$

Thus we have

$$\overline{a}(s)b(s)\overline{s}e^{-sR/v}$$
$$=\alpha\overline{a}(j\omega)b(j\omega)\cos kR - j\alpha\overline{a}(j\omega)b(j\omega)\sin kR$$
$$-j\omega\overline{a}(j\omega)b(j\omega)\cos kR - \omega\overline{a}(j\omega)b(j\omega)\sin kR$$
$$+j\omega\alpha\frac{R}{v}\overline{a}(j\omega)b(j\omega)\cos kR + \alpha\omega\frac{R}{v}\overline{a}(j\omega)b(j\omega)\sin kR \quad (25)$$
$$-j\omega\alpha\Omega_1(j\omega)\cos kR - \alpha\omega\Omega_1(j\omega)\sin kR$$
$$+\alpha\omega\Omega_2(j\omega)\cos kR - j\omega\alpha\Omega_2(j\omega)\sin kR + o(\alpha),$$

$$\overline{a}(s)b(s)se^{-sR/v}$$
$$=\alpha\overline{a}(j\omega)b(j\omega)\cos kR - j\alpha\overline{a}(j\omega)b(j\omega)\sin kR$$
$$+j\omega\overline{a}(j\omega)b(j\omega)\cos kR + \omega\overline{a}(j\omega)b(j\omega)\sin kR$$
$$-j\omega\alpha\frac{R}{c}\overline{a}(j\omega)b(j\omega)\cos kR - \alpha\omega\frac{R}{v}\overline{a}(j\omega)b(j\omega)\sin kR \quad (26)$$
$$+j\omega\alpha\Omega_1(j\omega)\cos kR + \alpha\omega\Omega_1(j\omega)\sin kR$$
$$-\alpha\omega\Omega_2(j\omega)\cos kR + j\omega\alpha\Omega_2(j\omega)\sin kR + o(\alpha).$$

Now all terms in (23) may be expanded into a Taylor series at $\alpha=0$. By comparing the coefficients of similar terms and making use of (23), (25) and (26), we immediately obtain (6), (7) and the following

$$\tilde{W}_m+\tilde{W}_e=\frac{\eta v}{16\pi}\iint_{VV}\frac{\overline{\rho}(\mathbf{r})\rho(\mathbf{r}')}{R}\cos kR\, dV(\mathbf{r})dV(\mathbf{r}')$$
$$+\frac{\eta v}{16\pi}\iint_{VV}\omega\frac{R}{v}\frac{\overline{\rho}(\mathbf{r})\rho(\mathbf{r}')}{R}\sin kR\, dV(\mathbf{r})dV(\mathbf{r}')$$
$$-\frac{\eta v}{16\pi}\iint_{VV}\frac{\omega\Omega_{\rho 1}(j\omega)}{R}\sin kR\, dV(\mathbf{r})dV(\mathbf{r}')$$
$$+\frac{\eta v}{16\pi}\iint_{VV}\frac{\omega\Omega_{\rho 2}(j\omega)}{R}\cos kR\, dV(\mathbf{r})dV(\mathbf{r}')$$
$$+\frac{\eta v}{16\pi}\iint_{VV}\frac{1}{v^2}\frac{\overline{\mathbf{J}}(\mathbf{r})\cdot \mathbf{J}(\mathbf{r}')}{R}\cos kR\, dV(\mathbf{r})dV(\mathbf{r}') \quad (27)$$
$$-\frac{\eta v}{16\pi}\iint_{VV}\frac{1}{v^2}\omega\frac{R}{v}\frac{\overline{\mathbf{J}}(\mathbf{r})\cdot \mathbf{J}(\mathbf{r}')}{R}\sin kR\, dV(\mathbf{r})dV(\mathbf{r}')$$
$$+\frac{\eta v}{16\pi}\iint_{VV}\frac{1}{v^2}\frac{\omega\Omega_{J1}(j\omega)}{R}\sin kR\, dV(\mathbf{r})dV(\mathbf{r}')$$
$$-\frac{\eta v}{16\pi}\iint_{VV}\frac{1}{v^2}\frac{\omega\Omega_{J2}(j\omega)}{R}\cos kR\, dV(\mathbf{r})dV(\mathbf{r}'),$$

where

$$\Omega_{\rho 1}(j\omega)=\operatorname{Re}\left[j\frac{\partial \overline{\rho}(\mathbf{r})}{\partial \omega}\rho(\mathbf{r}')-j\overline{\rho}(\mathbf{r})\frac{\partial \rho(\mathbf{r}')}{\partial \omega}\right],$$
$$\Omega_{\rho 2}(j\omega)=\operatorname{Im}\left[j\frac{\partial \overline{\rho}(\mathbf{r})}{\partial \omega}\rho(\mathbf{r}')-j\overline{\rho}(\mathbf{r})\frac{\partial \rho(\mathbf{r}')}{\partial \omega}\right].$$
$$\Omega_{J1}(j\omega)=\operatorname{Re}\left[j\frac{\partial \overline{\mathbf{J}}(\mathbf{r})}{\partial \omega}\cdot \mathbf{J}(\mathbf{r}')-j\overline{\mathbf{J}}(\mathbf{r})\cdot\frac{\partial \mathbf{J}(\mathbf{r}')}{\partial \omega}\right],$$
$$\Omega_{J2}(j\omega)=\operatorname{Im}\left[j\frac{\partial \overline{\mathbf{J}}(\mathbf{r})}{\partial \omega}\cdot \mathbf{J}(\mathbf{r}')-j\overline{\mathbf{J}}(\mathbf{r})\cdot\frac{\partial \mathbf{J}(\mathbf{r}')}{\partial \omega}\right].$$

Note that the integral

$$\iint_{VV}\left[j\frac{\partial \overline{\rho}(\mathbf{r})}{\partial \omega}\rho(\mathbf{r}')-j\overline{\rho}(\mathbf{r})\frac{\partial \rho(\mathbf{r}')}{\partial \omega}\right]\frac{\sin kR}{R}dV(\mathbf{r})dV(\mathbf{r}')$$

is real. Thus (27) can be written as

$$\tilde{W}_m + \tilde{W}_e =$$
$$\frac{\eta v}{16\pi} \iint_{V\,V} \left[ \bar{\rho}(\mathbf{r})\rho(\mathbf{r}') + \frac{1}{v^2} \bar{\mathbf{J}}(\mathbf{r}) \cdot \mathbf{J}(\mathbf{r}') \right] \frac{\cos kR}{R} dV(\mathbf{r})dV(\mathbf{r}')$$
$$+ \frac{\omega\eta}{16\pi} \iint_{V\,V} \left[ \bar{\rho}(\mathbf{r})\rho(\mathbf{r}') - \frac{1}{c^2} \bar{\mathbf{J}}(\mathbf{r}) \cdot \mathbf{J}(\mathbf{r}') \right] \sin kR\, dV(\mathbf{r})dV(\mathbf{r}') \quad (28)$$
$$- \frac{\omega\eta v}{8\pi} \iint_{V\,V} \mathrm{Im}\left[ \bar{\rho}(\mathbf{r}) \frac{\partial \rho(\mathbf{r}')}{\partial \omega} \right] \frac{\sin kR}{R} dV(\mathbf{r})dV(\mathbf{r}')$$
$$+ \frac{\omega\eta v}{8\pi} \iint_{V\,V} \frac{1}{v^2} \mathrm{Im}\left[ \bar{\mathbf{J}}(\mathbf{r}) \cdot \frac{\partial \mathbf{J}(\mathbf{r}')}{\partial \omega} \right] \frac{\sin kR}{R} dV(\mathbf{r})dV(\mathbf{r}').$$

From (7) and (28), the stored energies can be found as follows

$$\tilde{W}_m = \frac{\eta v}{16\pi} \iint_{V\,V} \frac{1}{v^2} \bar{\mathbf{J}}(\mathbf{r}) \cdot \mathbf{J}(\mathbf{r}') \frac{\cos kR}{R} dV(\mathbf{r})dV(\mathbf{r}')$$
$$+ \frac{\omega\eta}{32\pi} \iint_{V\,V} \left[ \bar{\rho}(\mathbf{r})\rho(\mathbf{r}') - \frac{1}{v^2} \bar{\mathbf{J}}(\mathbf{r}) \cdot \mathbf{J}(\mathbf{r}') \right] \sin kR\, dV(\mathbf{r})dV(\mathbf{r}')$$
$$- \frac{\omega\eta v}{16\pi} \iint_{V\,V} \mathrm{Im}\left[ \bar{\rho}(\mathbf{r}) \frac{\partial \rho(\mathbf{r}')}{\partial \omega} \right] \frac{\sin kR}{R} dV(\mathbf{r})dV(\mathbf{r}') \quad (29)$$
$$+ \frac{\omega\eta v}{16\pi} \iint_{V\,V} \frac{1}{v^2} \mathrm{Im}\left[ \bar{\mathbf{J}}(\mathbf{r}) \cdot \frac{\partial \mathbf{J}(\mathbf{r}')}{\partial \omega} \right] \frac{\sin kR}{R} dV(\mathbf{r})dV(\mathbf{r}'),$$

$$\tilde{W}_e = \frac{\eta v}{16\pi} \iint_{V\,V} \bar{\rho}(\mathbf{r})\rho(\mathbf{r}') \frac{\cos kR}{R} dV(\mathbf{r})dV(\mathbf{r}')$$
$$+ \frac{\omega\eta}{32\pi} \iint_{V\,V} \left[ \bar{\rho}(\mathbf{r})\rho(\mathbf{r}') - \frac{1}{v^2} \bar{\mathbf{J}}(\mathbf{r}) \cdot \mathbf{J}(\mathbf{r}') \right] \sin kR\, dV(\mathbf{r})dV(\mathbf{r}')$$
$$- \frac{\omega\eta v}{16\pi} \iint_{V\,V} \mathrm{Im}\left[ \bar{\rho}(\mathbf{r}) \frac{\partial \rho(\mathbf{r}')}{\partial \omega} \right] \frac{\sin kR}{R} dV(\mathbf{r})dV(\mathbf{r}') \quad (30)$$
$$+ \frac{\omega\eta v}{16\pi} \iint_{V\,V} \frac{1}{v^2} \mathrm{Im}\left[ \bar{\mathbf{J}}(\mathbf{r}) \cdot \frac{\partial \mathbf{J}(\mathbf{r}')}{\partial \omega} \right] \frac{\sin kR}{R} dV(\mathbf{r})dV(\mathbf{r}').$$

Equations (29) and (30) are the general expressions of the stored energies for a lossless antenna. The last two terms in (29) and (30) will be denoted by

$$\tilde{W}_d = -\frac{\omega\eta v}{16\pi} \iint_{V\,V} \mathrm{Im}\left[ \bar{\rho}(\mathbf{r}) \frac{\partial \rho(\mathbf{r}')}{\partial \omega} \right] \frac{\sin kR}{R} dV(\mathbf{r})dV(\mathbf{r}')$$
$$+ \frac{\omega\eta v}{16\pi} \iint_{V\,V} \frac{1}{v^2} \mathrm{Im}\left[ \bar{\mathbf{J}}(\mathbf{r}) \cdot \frac{\partial \mathbf{J}(\mathbf{r}')}{\partial \omega} \right] \frac{\sin kR}{R} dV(\mathbf{r})dV(\mathbf{r}'), \quad (31)$$

which represent frequency-derivative terms of the source distributions and disappear in (10) and (11) due to the incorrect assumption that the source distributions are independent of frequency. The contribution of $\tilde{W}_d$ to the stored energies could be significant, and cannot be ignored in general except for small antennas. Equations (29) and (30) reduce to (10) and (11) if either the source distributions are assumed to be independent of frequency or the source distributions are purely real (or imaginary). Also note that (29) and (30) reduce to (8) and (9) when the antenna becomes small. It should be notified that the stored energies given by (3), (4), (29) and (30) include the stored energies inside the circumscribing sphere of the antenna. Therefore the Q obtained is the actual value, not a lower bound. Numerical analysis indicates that the first term on the right-hand side of (29) or (30) is the dominant part for small antennas. This reveals that the stored energies around a small antenna are mainly quasi-static.

### III. MINIMIZATION OF ANTENNA Q

It is known that the antenna Q has a lower bound [1],[4],[20]. The devoted efforts of many researchers have been focused on the optimization of antenna geometries to approach the lower bound. The lower bound can also be approached by optimizing the current distribution for fixed antenna geometry. Making use of the continuity equation, we may rewrite (28) and (6) as

$$\tilde{W}_m + \tilde{W}_e =$$
$$\frac{\eta}{16\pi v} \iint_{V\,V} \left[ \bar{\mathbf{J}}(\mathbf{r}) \cdot \mathbf{J}(\mathbf{r}') \frac{\cos kR}{R} - \frac{1}{k^2} \mathbf{J}(\mathbf{r}') \cdot \nabla'\nabla' \frac{\cos kR}{R} \cdot \bar{\mathbf{J}}(\mathbf{r}) \right] dV(\mathbf{r})dV(\mathbf{r}')$$
$$+ \frac{k\eta}{16\pi v} \iint_{V\,V} \left[ \bar{\mathbf{J}}(\mathbf{r}) \cdot \mathbf{J}(\mathbf{r}') \sin kR + \frac{1}{k^2} \mathbf{J}(\mathbf{r}') \cdot \nabla'\nabla' \sin kR \cdot \bar{\mathbf{J}}(\mathbf{r}) \right] dV(\mathbf{r})dV(\mathbf{r}')$$
$$+ \frac{k\eta}{8\pi} \mathrm{Im} \iint_{V\,V} \frac{\partial \mathbf{J}(\mathbf{r}')}{\partial \omega} \cdot \left( \bar{\mathbf{I}} + \frac{1}{k^2}\nabla'\nabla' \right) \frac{\sin kR}{R} \cdot \bar{\mathbf{J}}(\mathbf{r}) dV(\mathbf{r})dV(\mathbf{r}'),$$
$$\quad (32)$$

$$P^{rad} = \frac{k\eta}{8\pi} \iint_{V\,V} \mathbf{J}(\mathbf{r}') \cdot \left( \bar{\mathbf{I}} + \frac{1}{k^2}\nabla'\nabla' \right) \frac{\sin kR}{R} \cdot \bar{\mathbf{J}}(\mathbf{r}) dV(\mathbf{r})dV(\mathbf{r}'). \quad (33)$$

Both the total stored energy and the radiated power can be represented as an inner product

$$\tilde{W}_m + \tilde{W}_e = (\hat{A}\mathbf{J}, \mathbf{J}), \quad P^{rad} = (\hat{B}\mathbf{J}, \mathbf{J}),$$

where

$$\hat{A}\mathbf{J}(\mathbf{r}) = \frac{\eta}{16\pi v} \int_V \left\{ \mathbf{J}(\mathbf{r}') \frac{\cos kR}{R} - \frac{1}{k^2}[\mathbf{J}(\mathbf{r}') \cdot \nabla']\nabla' \frac{\cos kR}{R} \right\} dV(\mathbf{r}')$$
$$+ \frac{\eta}{16\pi v} \int_V \left\{ \mathbf{J}(\mathbf{r}')k\sin kR + \frac{1}{k}[\mathbf{J}(\mathbf{r}') \cdot \nabla']\nabla' \sin kR \right\} dV(\mathbf{r}') \quad (34)$$
$$+ \frac{\eta}{8\pi v} \mathrm{Im} \int_V \left\{ \frac{k\partial \mathbf{J}(\mathbf{r}')}{\partial k} \frac{\sin kR}{R} + \frac{1}{k}\left[\frac{\partial \mathbf{J}(\mathbf{r}')}{\partial k} \cdot \nabla'\right]\nabla' \frac{\sin kR}{R} \right\} dV(\mathbf{r}'),$$

$$\hat{B}\mathbf{J} = \frac{k\eta}{8\pi} \int_V \left\{ \mathbf{J}(\mathbf{r}') + \frac{1}{k^2}[\mathbf{J}(\mathbf{r}') \cdot \nabla']\nabla' \right\} \frac{\sin kR}{R} dV(\mathbf{r}'). \quad (35)$$

Thus the radiation Q can be written as a Rayleigh quotient

$$Q = \frac{\omega(\tilde{W}_e + \tilde{W}_m)}{P^{rad}} = \frac{(\hat{A}\mathbf{J}, \mathbf{J})}{(\hat{B}\mathbf{J}, \mathbf{J})}. \quad (36)$$

Now we seek an optimal solution $\mathbf{J}$ that minimizes the Q in a permission region consisting of real vector functions. Under this condition, the operator $\hat{A}$ becomes a symmetric operator. By varational analysis, the optimal solution $\mathbf{J}$ must satisfy the following generalized eigenvalue equation

$$\hat{B}\mathbf{J} = \alpha \hat{A}\mathbf{J}, \quad (37)$$

where $\alpha = 1/Q$. The above approach is similar to the optimization of the ratio of gain to Q, which also needs to solve a generalized eigenvalue equation [1]. A major difference is that the generalized eigenvalue equation for the ratio of gain to Q has a unique solution, while the generalized eigenvalue equation (37) has more than one solution. The largest eigenvalue $\alpha$ gives the smallest value of the Q.

It is noted that a convex optimization procedure has been used by Gustafsson *et al* to determine the optimal current distributions that provide upper bounds on the antenna performance [35][36], and the procedure as well as the target functional are quite different from what is presented here.

## IV. NUMERICAL RESULTS

In this section, the new expressions (29) and (30) are applied to typical thin wire antennas to demonstrate their validity. The generalized eigenvalue equation (37) is used to determine the optimal current distributions for several typical thin wire geometries, which minimizes antenna Q.

### A. Calculations of stored energies and radiation Q

In order to find the stored energies around an antenna using (3) and (4) or (29) and (30), the induced current distribution $\mathbf{J}$ must be a known quantity. For convenience, the expressions (29) and (30) will be applied to thin wire antennas excited by a delta voltage source, for which the current distributions may be found by asymptotic analysis [22].

*1) Dipole antenna Q*

An arbitrary thin wire dipole antenna of length $L$ is shown in Figure 3. The wire is assumed to be a curved circular cylinder of radius $a_0$ and a curvilinear $l$-axis ($l$ stands for arc length) runs along the axis of the circular cylinder. The dipole is excited by delta voltage source at $l = l'$:

$$V = \mathbf{E}^{in} \cdot \mathbf{u}_l(l) = V_s \delta(l - l'). \tag{38}$$

where $\mathbf{E}^{in}$ is the incident field at $l = l'$ and $\mathbf{u}_l(l)$ is the unit tangent vector along $l$-axis.

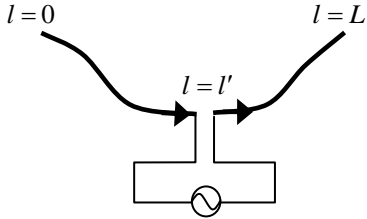

Figure 3 A dipole antenna excited by a delta voltage source.

It can be shown that the current distribution for the thin dipole antenna is given by [22]

$$I(l) = \frac{j\pi V_s}{\eta \ln ka_0 \sin kL}\left[\cos k\left(L - |l - l'|\right) - \cos k(L - l - l')\right]. \tag{39}$$

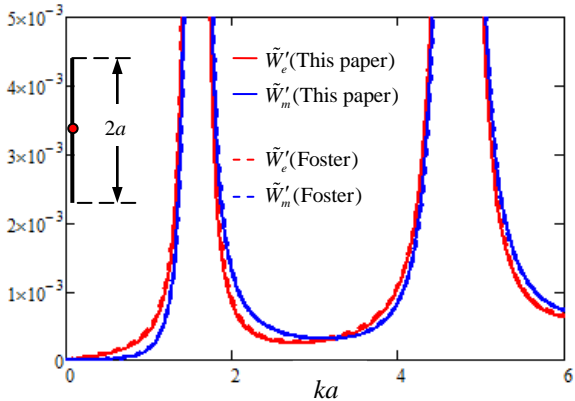

Figure 4 Stored energies of a straight thin wire dipole antenna ($\tilde{W}'_{e(m)} = \omega \tilde{W}_{e(m)} / |V_s|^2$).

For a straight thin wire dipole antenna centrally fed by delta voltage source with $a_0/a = 10^{-5}$, the stored energies of the dipole antenna are depicted in Figure 4. The solid lines represent the calculated results from (29) and (30), and the dashed lines are the results from the Foster reactance theorem (3) and (4). It can be seen that both methods give the same results. Figure 5 shows the quality factor Q obtained from the two methods, and a good agreement is obtained. For comparison, the Q obtained from the traditional definition is also shown in the same figure.

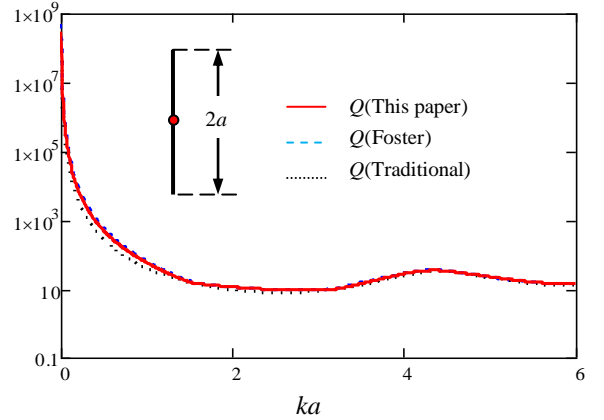

Figure 5 Q of thin wire dipole antenna.

*2) Loop antenna Q*

A thin-wire loop antenna of radius $a_0$ is shown in Figure 6. The loop is excited by delta voltage source at $l = l'$ as given by (38). Exact solutions for the current distribution exist for a thin circular loop antenna [9],[38]-[40]. For a general loop, the solution of the current distribution can be found by asymptotic analysis as follows [22]

$$I(l) = \frac{j\pi V_s}{\eta \ln ka_0} \frac{\cos\left(kL/2 - k|l - l'|\right)}{\sin(kL/2)}. \tag{40}$$

Setting $l' = L/2$, we have

$$I(l) = \frac{j\pi V_s}{\eta \ln ka_0} \begin{cases} \cos kl, & l < L/2 \\ \cos k(L - l), & l > L/2 \end{cases}.$$

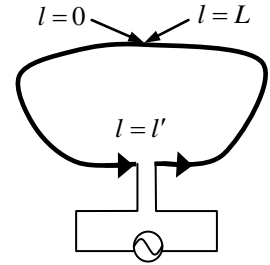

Figure 6 A loop antenna excited by a delta voltage source.

The stored energies and the frequency-derivative term $\tilde{W}_d$ defined in (31) for a thin circular loop with $a_0/a = 10^{-2}$ are plotted in Figure 7, which indicates that the contribution from $\tilde{W}_d$ is quite significant and cannot be ignored except for small antennas. In the calculation, the exact solution for the current distribution of the thin circular loop antenna has been used, which can be found in [9].

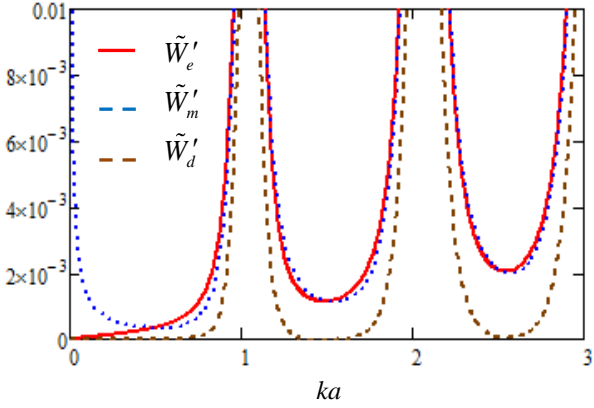

Figure 7 Stored energies ( $\tilde{W}'_{e(m)} = \omega \tilde{W}_{e(m)} / |V_s|^2$ )
and $\tilde{W}'_d = \omega \tilde{W}_d / |V_s|^2$.

The stored energies of thin wire loop antenna with $a_0/a = 10^{-5}$ have also been evaluated by using (3), (4), (29),(30) and (40). The numerical results are shown in Figure 8 and Figure 9. Again both methods produce the same results.

Note that the current distributions (39) and (40) are dependent of frequency.

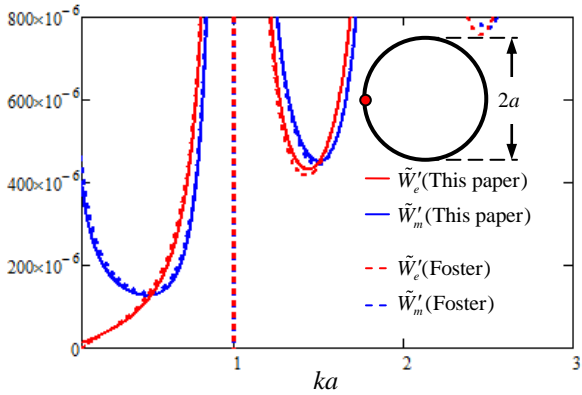

Figure 8 Stored energies of thin wire loop antenna
( $\tilde{W}'_{e(m)} = \omega \tilde{W}_{e(m)} / |V_s|^2$ ).

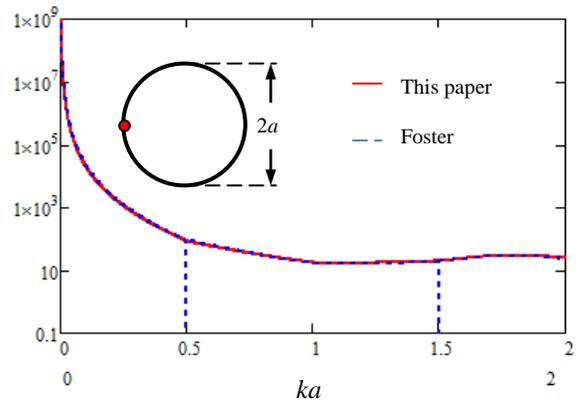

Figure 9 Q of thin wire loop antenna.

### B. Minimization of Q

When the antenna geometry is specified, (37) can be used to find the optimal current distribution that yields the minimum possible Q for the specified antenna geometry. In what follows, the maximum size (i.e., the diameter of the circumscribing sphere) of all antenna geometries examined is assumed to be $2a = 0.15\lambda$, where $\lambda$ is the wavelength, and the radius of the wire is assumed to be $a_0 = 10^{-5}\lambda$. The optimal current distributions along the length variable $s$ for several typical wire geometries are shown in Figure 10 to Figure 13. The corresponding Q values are listed in TABLE I. It can be seen that the Q value decreases as the antenna structure gets more complicated and occupies space more efficiently.

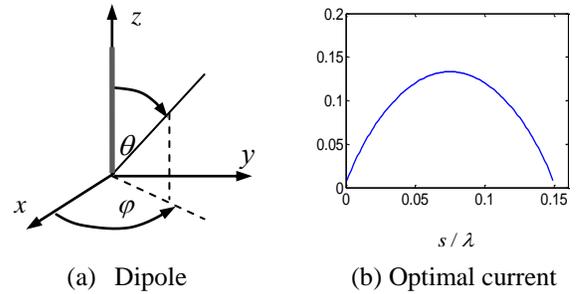

(a) Dipole   (b) Optimal current

Figure 10 The optimal current distribution of dipole antenna

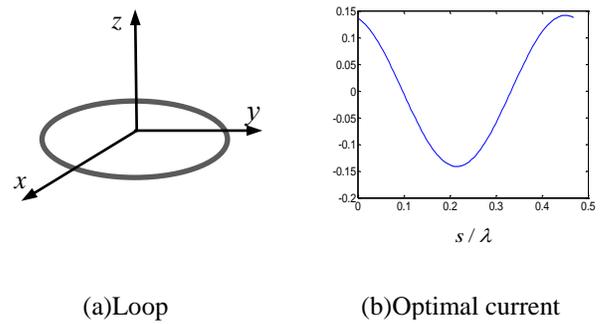

(a) Loop   (b) Optimal current

Figure 11 The optimal current distribution of loop antenna

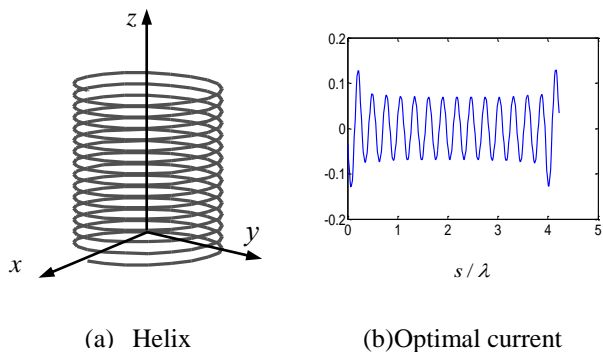

(a) Helix     (b) Optimal current

Figure 12 The optimal current distribution of helix antenna.

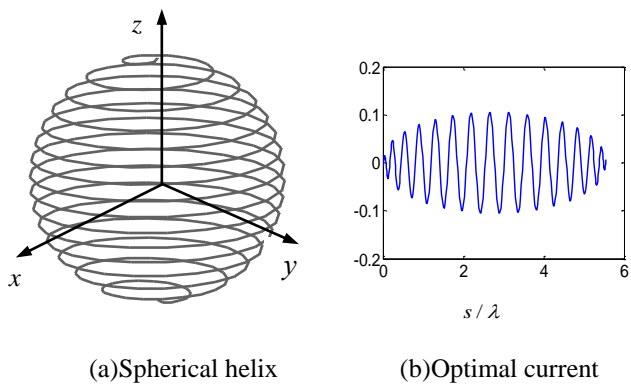

(a) Spherical helix     (b) Optimal current

Figure 13 Optimal current distribution of spherical helix antenna

TABLE I Optimized Q (maximum antenna size: $2a = 0.15\lambda$)

| Antenna | Dipole | Loop | Helix | Spherical helix |
|---------|--------|------|-------|-----------------|
| Q       | 125.99 | 70.98 | 35.71 | 15.75 |

## V. CONCLUSION

Various methods for evaluating the stored electromagnetic energies and radiation Q have been discussed in this paper. New expressions for the stored energies are obtained from the Poynting theorem in complex frequency domain, and they are compared with previous theory and are validated by numerical examples. The minimization of antenna Q is also investigated. It has been shown that an optimal current distribution that minimizes the Q for specified antenna geometry does exist.

It should be emphasized that the Foster reactance theorem (3) and (4) and the new expressions (29) and (30) obtained in this paper are independent and both are applicable to an arbitrary lossless antenna. A major difficulty in applying (3) and (4) is that the reactance curve of a lossless antenna may have singularities where the reactance approaches to infinity [13]. Many numerical methods fail to give accurate results for the reactance around the singularities, and thus may cause big numerical errors in the calculation of the frequency derivative of reactance (and thus stored energies and Q) due to its frequency sensitivity. This phenomenon is also observed by Harrington when he applies the Foster reactance theorem to the optimization of the ratio of gain to Q for antenna arrays [9]. Harrington points out that the frequency-sensitivity effects of the frequency derivative of antenna reactance become dominant so rapidly that little improvement of the gain-bandwidth product can be expected. For this reason, the direct computation of the frequency derivative of the reactance should be avoided when (3) and (4) are used. Instead, one can combine (3), (4), (5) and (7), and apply the frequency derivative to the integrand of (7). This process may reduce the numerical errors and improve the accuracy.

Finally we remark that our study is based on an assumption that the antenna and the surrounding medium are lossless. In 1964[41], Ginzburg pointed out that "Despite the fact that the problem of the conservation law and the expression for the energy density in electrodynamics is a fundamental one, there are certain aspects of it which have not yet been elucidated, in particular for the case of an absorbing dispersive medium." He also pointed out "When absorption is present, it is not in general possible to introduce phenomenologically the concept of the mean electromagnetic energy density." Ginzburg's above observations have revealed some challenges in fundamental electromagnetics. It is noted that one possible form of the average energy density applicable to dispersive and absorptive materials has been derived by Ruppin in 2002, and it remains positive at all frequencies[42].

The theory presented in this paper is also applicable to an antenna with lossless dielectric materials. According to compensation theorem for electromagnetic fields, the influence of the dielectric materials on the fields may be substituted by an equivalent source [6]. As long as the equivalent source is determined, (29) and (30) can be used to find the stored energies with the equivalent source included as part of the source distributions.

## APPENDIX

It has been proved in [5] that the Foster reactance theorem (3) and (4), which is well-known in circuit theory, is still valid for an ideal metal antenna subject to the conditions: (a) The antenna consists of perfect conductor and the surrounding material is lossless; (b) The antenna is fed by a waveguide which is assumed to be in a state of single-mode operation, and the antenna input terminal is positioned in the single mode region of the feeding waveguide. Equations (3) and (4) imply

$$\partial X / \partial \omega = 4(\tilde{W}_m + \tilde{W}_e) / |I|^2 > 0,$$

which indicates that the slope of the reactance of the antenna must be great than zero. The above result is the main issue to cause the argument [10]-[13].

A number of numerical examples will be presented in this Appendix to further validate the Foster reactance theorem for ideal (lossless) antennas. For other numerical examples and related discussion, please refer to [5], [9]-[13].

### A. Antennas fed by transmission lines

The reactance curves of a monopole antenna fed by a coaxial cable and rectangular aperture antenna fed by a waveguide are shown in Figure 14 and Figure 15. It can be seen that Foster reactance theorem holds very well in the frequency range between the cut-off frequency $f_c$ of the dominant mode and the cut-off frequency $f_{c1}$ of the first higher order mode (the feeding waveguide is assumed to be in a single-mode operation).

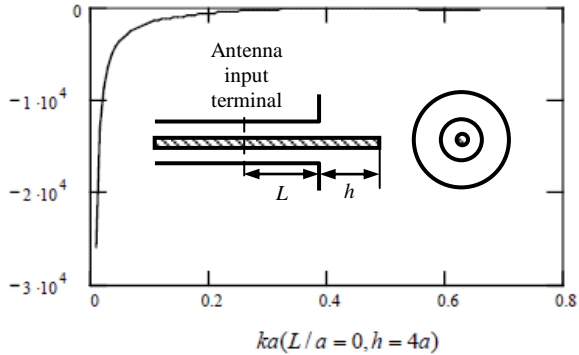

Figure 14 Reactance of a cylindrical monopole antenna.

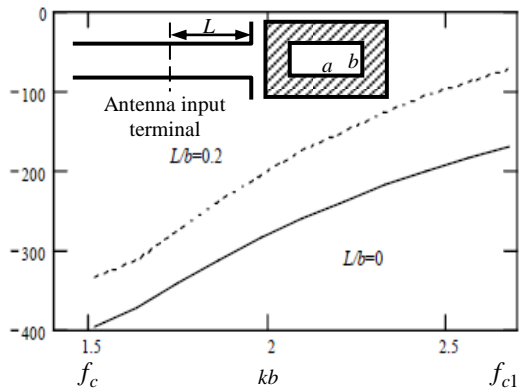

Figure 15 Reactance of a rectangular aperture.

### B. Point-fed antennas

Although a feeding waveguide is assumed in the proof of the Foster reactance theorem [5], (3) and (4) are also valid for point-fed antennas. As pointed out by Balanis[43], the delta-gap source modeling is the simplest and most widely used, but it is also the least accurate, especially for impedances. Theoretically the delta-gap source is only valid for infinitely thin wire antennas. For thick wires, the delta-gap source modeling produces reasonable results for the impedance only when the calculation is limited to the low frequency range. This observation has been widely ignored.

The reactance curves of the thin dipole antenna, loop antenna and folded dipole antenna are shown in Figure 16, Figure 17 and Figure 18. It can be seen that the Foster reactance theorem holds very well. Note that the reactance curves not only have some zeros (resonant frequencies) where the stored electric energy equals the stored magnetic energy, but also have some singularities where the slope of the reactance curve becomes infinite. A negative slope may occur around these singularities if either the delta-gap source is inappropriately applied or heat loss is present (such as in the measurements). However the slope of reactance is always positive in the vicinity of resonant frequency as noted by Rhodes [11]. Since the bandwidth and Q are defined at resonant frequency, the Foster reactance theorem (3) and (4) are always approximately valid for a lossy system and can be applied to study antenna Q and bandwidth, as claimed by Harrington [9].

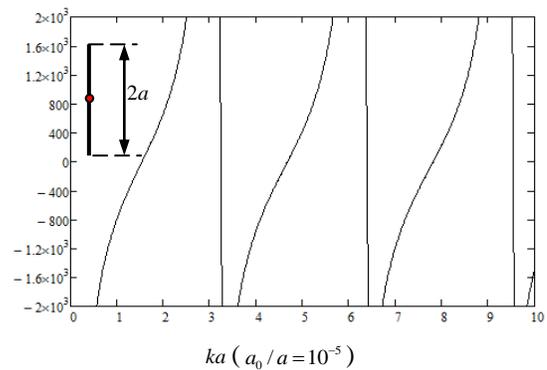

Figure 16 Reactance of a dipole antenna.

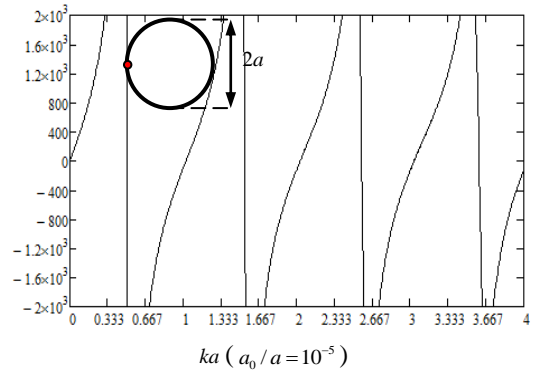

Figure 17 Reactance of a loop antenna.

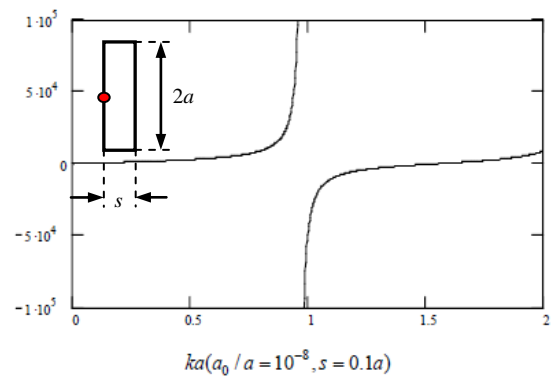

Figure 18 Reactance of a folded dipole.